\documentclass[prl,twocolumn,superscriptaddress]{revtex4-1}
\usepackage{graphicx}
\usepackage{amsmath}
\usepackage{amssymb}
\usepackage{siunitx}
\usepackage{mathrsfs}
\usepackage{xcolor}
\usepackage{multirow}
\usepackage[normalem]{ulem}
\usepackage{color}
\usepackage{chemformula} 
\usepackage{cancel}
\usepackage{graphicx}
\usepackage{dcolumn}
\usepackage{bm}
\usepackage{hyperref}
\definecolor{darkgreen}{rgb}{0.0, 0.5, 0.0}

\begin{document}

\preprint{APS/123-QED}

\title{On the Slipperiness of Surfactants: Charge-Mediated Friction Control at the Molecular Scale}%


\author{Kaili Xie}
\affiliation{Van der Waals-Zeeman Institute, Institute of Physics, University of Amsterdam, 1098XH, Amsterdam, The Netherlands}%

\author{Julie Jagielka}%
\affiliation{Van der Waals-Zeeman Institute, Institute of Physics, University of Amsterdam, 1098XH, Amsterdam, The Netherlands}%

\author{Liang Peng}%
\affiliation{Van der Waals-Zeeman Institute, Institute of Physics, University of Amsterdam, 1098XH, Amsterdam, The Netherlands}%

\author{Yu Han}%
\affiliation{Department of Molecular Spectroscopy, Max Planck Institute for Polymer Research, 55128, Mainz, Germany}%

\author{Yedam Lee}%
\affiliation{Department of Molecular Spectroscopy, Max Planck Institute for Polymer Research, 55128, Mainz, Germany}%

\author{Steve Franklin}%
\affiliation{Van der Waals-Zeeman Institute, Institute of Physics, University of Amsterdam, 1098XH, Amsterdam, The Netherlands}%
\affiliation{Advanced Research Center for Nanolithography, 1098XG, Amsterdam, The Netherlands}

\author{Yongkang Wang}%
\affiliation{Department of Molecular Spectroscopy, Max Planck Institute for Polymer Research, 55128, Mainz, Germany}%

\author{Arsh Hazrah}%
\affiliation{Department of Molecular Spectroscopy, Max Planck Institute for Polymer Research, 55128, Mainz, Germany}%

\author{Mischa Bonn}%
\affiliation{Van der Waals-Zeeman Institute, Institute of Physics, University of Amsterdam, 1098XH, Amsterdam, The Netherlands}%
\affiliation{Department of Molecular Spectroscopy, Max Planck Institute for Polymer Research, 55128, Mainz, Germany}%

\author{Joshua Dijksman}%
\affiliation{Van der Waals-Zeeman Institute, Institute of Physics, University of Amsterdam, 1098XH, Amsterdam, The Netherlands}

\author{Daniel Bonn}%
\email{d.bonn@uva.nl}
\affiliation{Van der Waals-Zeeman Institute, Institute of Physics, University of Amsterdam, 1098XH, Amsterdam, The Netherlands}%



\date{\today}

\begin{abstract}
From soap-covered dishes to freshly cleaned floors, surfactants can make surfaces slippery; yet, the underlying mechanism remains poorly understood. Here, we identify the molecular origin behind this ubiquitous phenomenon using macroscopic tribology and surface molecular spectroscopy. We demonstrate that reducing friction through surfactants hinges on charge complementarity: surfactants of opposite charge to the solid surface reduce friction even at extreme contact pressures, whereas like-charged or neutral surfactants are ineffective. Oppositely charged surfactants self-assemble into dense and robust molecular brushes, creating a persistent lubrication beyond the limits of conventional mechanisms. This charge-mediated approach offers a universal and scalable framework for friction control across length scales without significant surface modification.

\end{abstract}



\maketitle



\textit{Introduction---}The impact of friction and slipperiness spans scales from molecular motors~\cite{fisher1999force,iino2020introduction,schliwa2003molecular} to industrial tribology~\cite{bowden2001friction,holmberg2012global,mang2011industrial}. Surfactants are known to ubiquitously reduce friction in aqueous environments—facilitating processes from biological lubrication~\cite{qin2025tribology} to everyday cleaning~\cite{kim2013functional}—yet the molecular origins responsible for this behavior have proven elusive, particularly under the extreme conditions where conventional hydrodynamic lubrication fails.

The primary challenge arises in the boundary lubrication regime~\cite{persson1993theory,briscoe2006boundary,briscoe2017aqueous,peng2022nonmonotonic}, where the sliding surfaces are in direct contact through asperities. In this regime, when the lubricant film thickness approaches molecular dimensions, the properties of individual molecules and their interactions with surfaces become the dominant factors governing friction~\cite{israelachvili2011intermolecular,raviv2003lubrication,toyoda2025boundary}. 
Recent experiments using atomic force microscopy (AFM) and surface force balance (SFB) have shown that coating surfaces with molecular layers, such as polymers~\cite{raviv2003lubrication,landherr2011interfacial}, polysaccharides~\cite{bresme2022electrotunable}, lipids~\cite{lin2020cartilage}, and surfactants~\cite{kamada2011surfactant,silbert2014normal}, results in low friction because they separate the solid surfaces from each other.
The lubrication mechanisms most commonly invoked in such cases—sacrificial molecular layers~\cite{chan2012tribological}, easily sheared interfacial films~\cite{briscoe2006boundary,toyoda2025boundary}, or micelle-based nanoscale 'ball bearing' effects~\cite{li2018improvement,vakarelski2004lateral,manne1995molecular}—are primarily based on the specific scenario in which the molecular layers remain intact during surface sliding. These mechanisms fail to account for the persistent slipperiness observed under high contact pressures~\cite{carpick1997scratching,fukagai2022transition}, where such supramolecular structures should collapse~\cite{silbert2014normal}. Once disrupted, the interfacial structures rarely self-repair and fail to maintain effective lubrication~\cite{raviv2003lubrication}.

In this Letter, we identify a distinct and robust mechanism for surfactant-governed lubrication: charge-mediated molecular adsorption on solid surfaces. We show that oppositely charged surfactant–surface pairs achieve dramatically lower friction than neutral or like-charged combinations. 
Surface spectroscopy provides direct molecular-level evidence that the specific adsorption of charged surfactant headgroups onto oppositely charged surface sites stabilizes the interfacial layer. Densely packed molecules enhance intermolecular interactions, rendering the surfactant layer robust against collapse even under extreme shear and compression.
This charge-dependent adsorption on surfaces offers a molecular-level explanation for the durability of surfactant-induced slipperiness and establishes principles for rational lubrication system design from synovial fluids~\cite{waller2013role} to industrial lubricants~\cite{mang2007lubricants}.

\begin{figure*}[ht]
    \centering
    \includegraphics[scale=1.25]{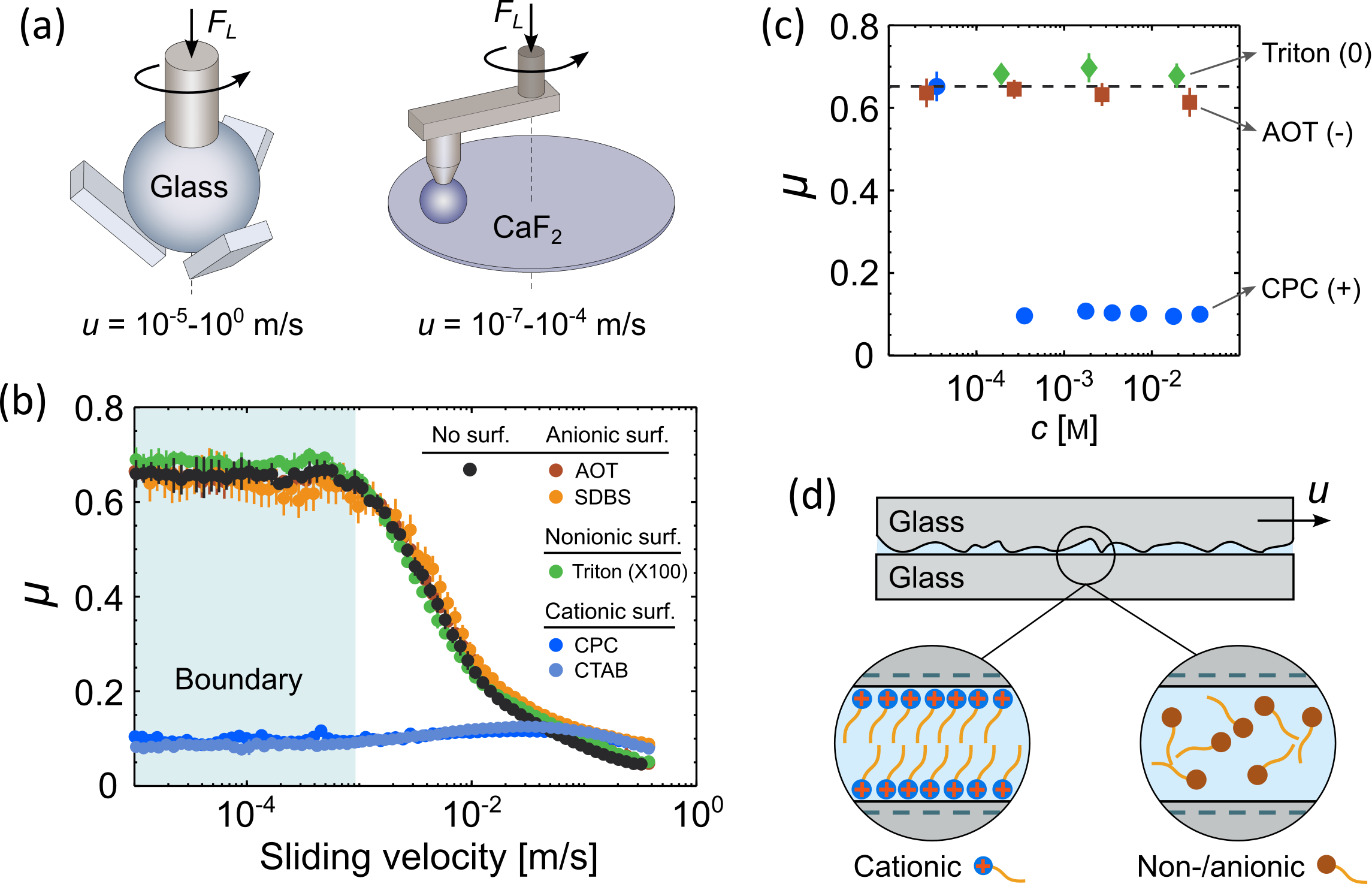}
    \caption{Tribological experiment. (a) Schematics of the friction experiment setups. A glass ball (left) slides over three glass plates inclined at an angle of $\theta$ = 1/4$\pi$, while a \ch{CaF2} sphere (right) slides over a horizontal \ch{CaF2} plate. The sliding velocity $u$ is carefully controlled, while the total normal load $F_L$ is maintained by an internal feedback loop. All sliding interfaces are fully immersed in a liquid environment. (b) Coefficient of friction $\mu$ between sliding glass surfaces versus sliding velocity in the presence of various surfactants. The boundary lubrication regime (shaded in blue) is identified by the plateau observed in the control experiment (no surfactant). All surfactants have a concentration of 10 mM. Error bars indicate the standard deviation over at least three independent measurements. (c) Coefficient of friction versus surfactant bulk concentration $c$ for the glass sphere on glass surfaces in the presence of various surfactants. The dashed line indicates the plateau regime value of $\mu$ for the control experiment. (d) Schematic of surfactant adsorption at a lubricated glass-glass interface. }
    \label{fig:setup}
\end{figure*}

\textit{Experiments---}In our experiments, we measure the friction between sliding surfaces immersed in an aqueous environment using both a commercial tribometer (Anton Paar MCR302 with a tribology cell T-PTD 200) and our custom-built setup described earlier \cite{peng2022nonmonotonic} (see Fig.~\ref{fig:setup}(a) and Supplemental Material, Fig.~S1~\cite{SuppMat}). In both systems, a sphere slides over a flat substrate under controlled normal load ($F_L$) and velocity ($u$). The total normal load is kept constant while the sliding velocity is varied over multiple orders of magnitude. The tribometer operates over a velocity range of 10$^{-5}$ -- 10$^0$ m/s,  whereas the custom-built setup covers 10$^{-7}$ -- 10$^{-4}$ m/s, complementing each other.
The upper spheres are much rougher than the corresponding bottom substrates (Fig.~S2). The coefficient of friction ($\mu$) between the sliding surfaces is determined from the torque ($M$) measured by the rheometer head~\cite{SuppMat}. To clearly resolve the boundary lubrication regime, a glycerol–water mixture (75 wt\% glycerol) is used as the solvent in all tribological experiments (see Fig.~S3). Various types and concentrations of surfactants, as well as different contact pressures, are systematically explored in our experiments. We note that the surfactants do not significantly affect the solvent viscosity, as only a small amount is added (Fig.~S5).

\textit{Results and discussion---}We first determine the kinetic friction between the glass surfaces. We use the reference liquid without adding surfactants as the control experiment. Fig.~\ref{fig:setup}(b) shows the Stribeck curve \cite{stribeck1902wesentlichen,veltkamp2021lubricated}, which relates friction to sliding velocity.
In the control experiment, across a sliding velocity range of 10$^{-5}$ to 10$^{-3}$ m/s, $\mu$ remains approximately constant at 0.66 $\pm$ 0.04, showing no dependence on the velocity. 
This plateau indicates that lubrication is predominantly governed by the boundary regime~\cite{briscoe2017aqueous}.
At higher sliding velocities, $\mu$ decreases monotonically, reflecting a transition to the mixed lubrication regime. This progressive reduction in friction is primarily attributed to the emergence of hydrodynamic lift~\cite{saintyves2016self}, which partially reduces the effective contact area between surface asperities. Here, we focus primarily on the boundary lubrication regime.

In contrast, we find that introducing cationic surfactants, i.e., cetylpyridinium chloride (CPC) and cetyltrimethylammonium bromide (CTAB), into the solution significantly reduces $\mu$ to below 0.1 in the boundary regime, an approximately 85\% decrease relative to the control experiment. In both cases, $\mu$ is only weakly dependent on the sliding velocity across the boundary and mixed regimes ($u > 10^{-3}$ m/s). However, adding anionic surfactants, i.e., bis(2-ethylhexyl) sulfosuccinate sodium salt (AOT) and sodium dodecylbenzenesulfonate (SDBS), as well as a non-ionic surfactant (Triton X-100), does not alter the friction curve compared to the control experiment, indicating negligible lubrication. 

To understand this observation, we first consider that amphiphilic surfactants form aggregates in aqueous solution due to the hydrophobicity of their alkyl chains~\cite{deGennes2003capillarity} when above the critical micelle concentration (CMC). Such self-assembled structures may separate the solid–solid contact and form a sliding interfacial layer that reduces friction~\cite{silbert2014normal,patrick1999surface,brick2007self}. In our case, however, increasing the concentration of anionic, cationic, or non-ionic surfactants beyond their CMC does not affect the respective coefficients of friction (Fig.~\ref{fig:setup}(c)). In addition, the typical contact pressures in our experiments are incompatible with self-assembled aggregates: pressure reaches values of $P = F_n/A_r \sim 100$ MPa, where $F_n \approx$ 800 mN is the normal force and $A_r \approx \SI{3000}{\micro\meter^2}$ is the estimated contact area determined using the method described in Ref.~\cite{weber2018molecular}. Local pressures can even reach the order of gigapascal (GPa) due to the surface asperities. These pressures, which are beyond the range typically accessible by AFM or SFB techniques, by far exceed the stability limit of the self-assembled structures under compression, typically $\sim$1 MPa \cite{briscoe2006boundary,silbert2014normal,vakarelski2004lateral}. The self-assembled aggregates, if formed, likely collapse and fail to lubricate, ruling out a mechanism based on such supramolecular structures.

Instead, we propose a charge-mediated molecular mechanism in which cationic surfactants adsorbed on the solid surface reduce friction. In a neutral aqueous environment, glass surfaces acquire a negative charge due to deprotonation of surface silanol (-\ch{SiOH}) groups. Consequently, the positively charged headgroups of cationic surfactants preferentially adsorb via electrostatic interactions, forming a more brush-like, ordered layer than anionic or nonionic surfactants~\cite{carpick1997scratching,lu2022ultralow}, as illustrated in Fig.~\ref{fig:setup}(d).



\begin{figure}[ht]
    \centering
    \includegraphics[scale=1.15]{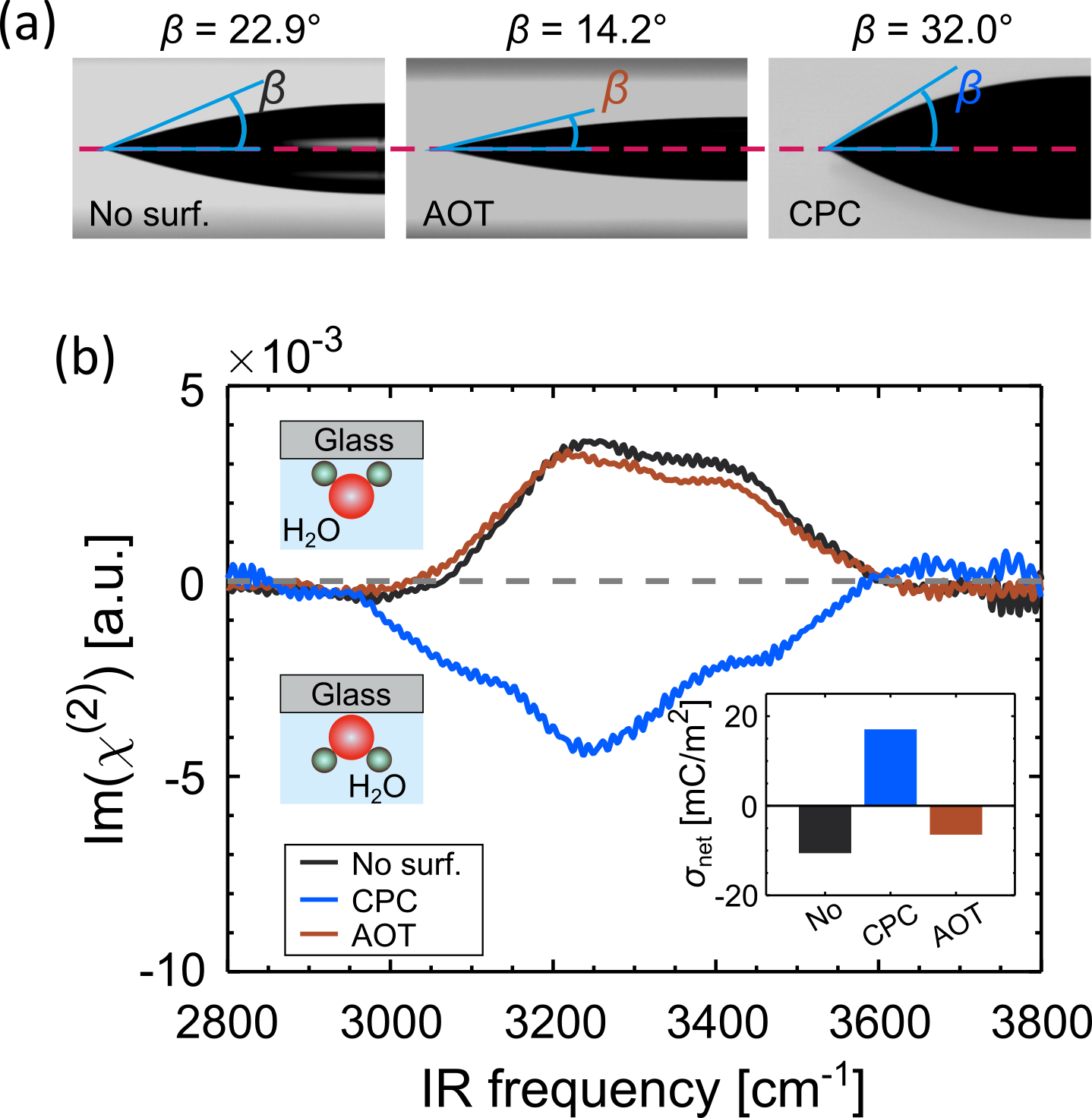}
    \caption{Characterization of surfactant adsorption on glass surface. (a) Contact angles of surfactant drops. The anionic surfactant (AOT) reduces the contact angle, whereas the cationic surfactant (CPC) increases it. (b) Im$\chi^{(2)}$ spectra of different surfactant solutions at the glass–water interface revealed by HD-SFG spectroscopy. The sign reversal between AOT and CPC highlights the critical role of charge-mediated surfactant adsorption. The surfactant solution has a concentration of 0.01 wt\% and contains 5 mM NaCl (see details in Supplemental Material). Inset cartoons (left) depict the orientation of water molecules at the glass surface. The inset plot (right) shows the estimated net surface charge density, $\sigma_{\mathrm{net}}$. The gray dashed line indicates zero. }
    \label{fig:SFG}
\end{figure}

To demonstrate such a mechanism, we first establish that the cationic surfactants indeed adsorb on the glass surface by measuring the contact angle $\beta$. As shown in Fig.~\ref{fig:SFG}(a), compared to a droplet without surfactants, a droplet with a cationic surfactant (CPC) exhibits a larger $\beta$ and a droplet with an anionic surfactant (AOT) exhibits a smaller $\beta$. This autophobic behavior~\cite{bera2021antisurfactant} of the cationic surfactant is then due to the oppositely charged molecules anchoring their headgroups to the surface, with their hydrophobic tails extending toward the air, leading to a large contact angle. We further confirm the surfactant adsorption on the glass surface by UV–Vis spectroscopy (Fig.~S6).




Quantitatively, we probe the charge interactions between surfactants and glass surface using heterodyne-detected sum-frequency generation (HD-SFG) spectroscopy~\cite{SuppMat}. HD-SFG is a surface-specific vibrational spectroscopy technique that selectively probes interfacial structures with molecular specificity~\cite{bonn2015molecular}. By targeting the O-H stretching mode of water, HD-SFG yields the complex second-order susceptibility spectrum ($\chi^{(2)}$) of the interfacial water molecules, revealing their absolute orientation and hydrogen-bonding network~\cite{nihonyanagi2009direct}, both of which are sensitive to the surface (dis-)charging~\cite{wen2016unveiling}, including that induced by surfactant adsorption~\cite{bera2021antisurfactant}. 
Fig.~\ref{fig:SFG}(b) shows HD-SFG spectra of the glass surface with and without surfactants. In the absence of surfactant, glass surface exhibits a broad positive band at 3000 – 3600 cm$^{-1}$, consistent with interfacial water molecules oriented with their hydrogens toward the negatively charged surface (net surface charge density of -10.6 mC/m$^2$, inferred by comparing the HD-SFG signals at different ionic strengths, following previously established protocols~\cite{wen2016unveiling,seki2021direct,wang2023chemistry}, top-left inset in Fig.~\ref{fig:SFG}(b)). Adding anionic surfactant AOT leaves the spectral feature largely unchanged, whereas the cationic surfactant CPC reverses its sign, indicating flipped water orientation and a reversal of the net surface charge to +17 mC/m$^2$ (insets in Fig.~\ref{fig:SFG}(b)).
Our HD-SFG measurements provide clear evidence that the cationic surfactant chemically adsorbs onto the glass surface, even inverting its surface charge, whereas the anionic surfactant does not.

\begin{figure}[ht]
    \centering
    \includegraphics[scale=1.13]{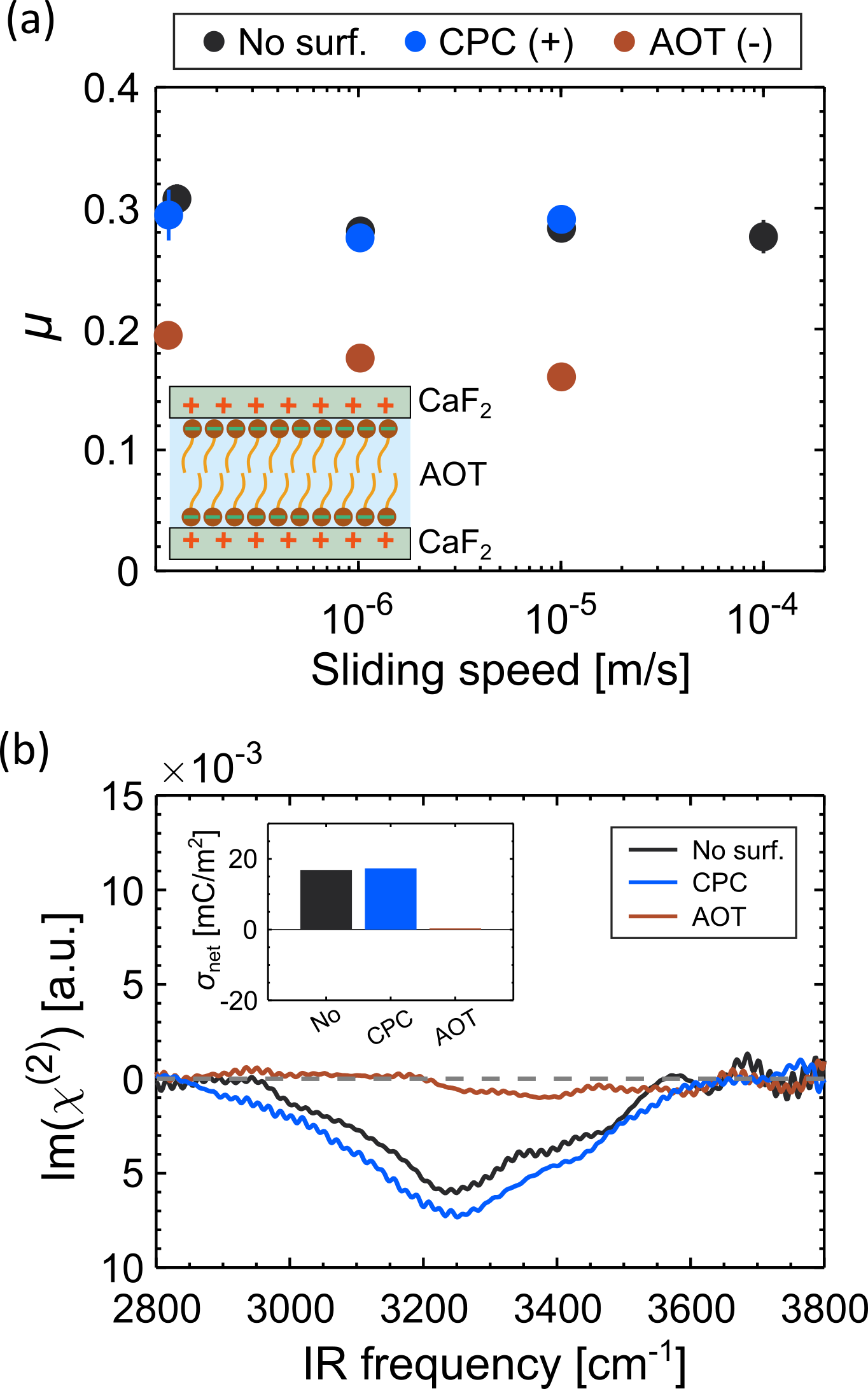}
    \caption{Friction and charge density at \ch{CaF2}-\ch{CaF2} interfaces. (a) Coefficient of friction $\mu$ versus sliding velocity. Different from glass surfaces, anionic surfactant AOT enhances the lubrication of the \ch{CaF2} interface, whereas the cationic surfactant CPC has a negligible effect. The friction measurement was performed using our customized setup (Fig.~\ref{fig:setup}(a)-right). The inset illustrates the adsorption of AOT on the surface. (b) HD-SFG spectra of different surfactants at the \ch{CaF2}-water interface. The inset shows the estimated net surface charge density. The gray dashed line indicates zero.}
    \label{fig:CaF2}
\end{figure}

To further study the interplay between surfactants and surfaces by charge, we investigate calcium fluoride (\ch{CaF2}) surfaces, which become positively charged in an aqueous environment at neutral pH~\cite{assemi2006isoelectric} as opposed to glass, which is negatively charged. We measure the friction using our custom-built setup (Fig.~\ref{fig:setup}(a)-right and Fig.~S1(b)). 
As shown in Fig.~\ref{fig:CaF2}(a), unlike the case of glass–glass friction, the anionic surfactant AOT markedly reduces friction between \ch{CaF2} surfaces, whereas the cationic CPC is ineffective. HD-SFG spectra confirm preferential AOT adsorption on \ch{CaF2}, shifting the net surface charge from positive toward negative, while CPC shows minimal adsorption (Fig.~\ref{fig:CaF2}(b)). Note that the friction reduction by AOT on \ch{CaF2} is weaker than that by CPC on glass, likely due to differences in the initial surface roughness and charge~\cite{raviv2003lubrication} as well as the surfactant type. Overall, the \ch{CaF2} results reinforce the picture of opposite charge-mediated surfactant adsorption and lubrication in the boundary regime.

\begin{figure}[ht]
    \centering
    \includegraphics[scale=1.13]{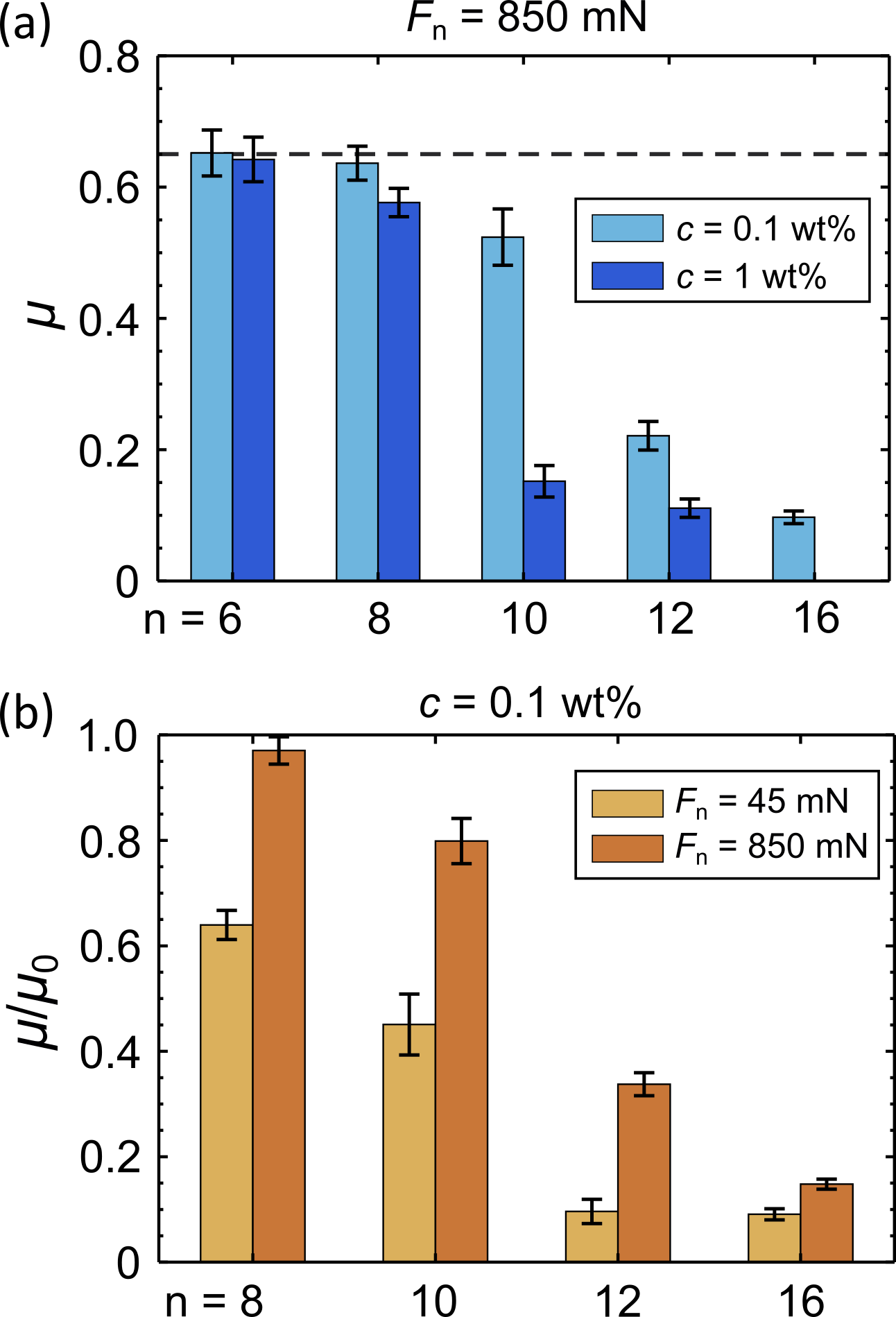}
    \caption{Effect of alkyl chain lengths on glass-glass friction coefficient $\mu$ for different C$_n$TAB concentrations $c$ (a) and normal forces $F_n$ (b). The data are average values from the plateau in the boundary lubrication regime (\textit{cf.} Fig.~\ref{fig:setup}(b)). The C$_n$TAB surfactants have different tail carbon numbers $n$ = 6, 8, 10, 12, and 16. The dashed line in (a) and $\mu_0$ in (b) indicate the friction coefficient without surfactants.  The normal force $F_n$ is the force applied perpendicularly to each glass surface; on average, $F_n = F_L/(3 \cos{\theta})$. }
    \label{fig:carbon_chain}
\end{figure}

We next test the charge-mediated mechanism by tailoring the lubrication of glass surfaces using cationic n-alkyltrimethylammonium bromide surfactants (C$_n$TAB) with different alkyl chain lengths, $n$ = 6 – 16. Fig.~\ref{fig:carbon_chain} shows that $\mu$ generally decreases with increasing alkyl chain length and is further governed by the initial surfactant concentration and applied compression. For example, under the normal force $F_n$ = 850 mN, surfactants with chain lengths $n \leq 8$ fail to lubricate effectively at any concentration, whereas longer chains yield a pronounced friction reduction with increasing concentration (Fig.~\ref{fig:carbon_chain}(a)). Under a lower compression (i.e., $F_n$ = 45 mN), as shown in Fig.~\ref{fig:carbon_chain}(b), chain lengths with $n$ = 8 - 16 exhibit relatively reduced friction, as surface asperities penetrate less into the surfactant film~\cite{carpick1997scratching}. These results suggest that effective boundary lubrication can be achieved by adsorbing long-chain surfactants onto the contact areas.

\begin{figure}[ht]
    \centering
    \includegraphics[scale=1.2]{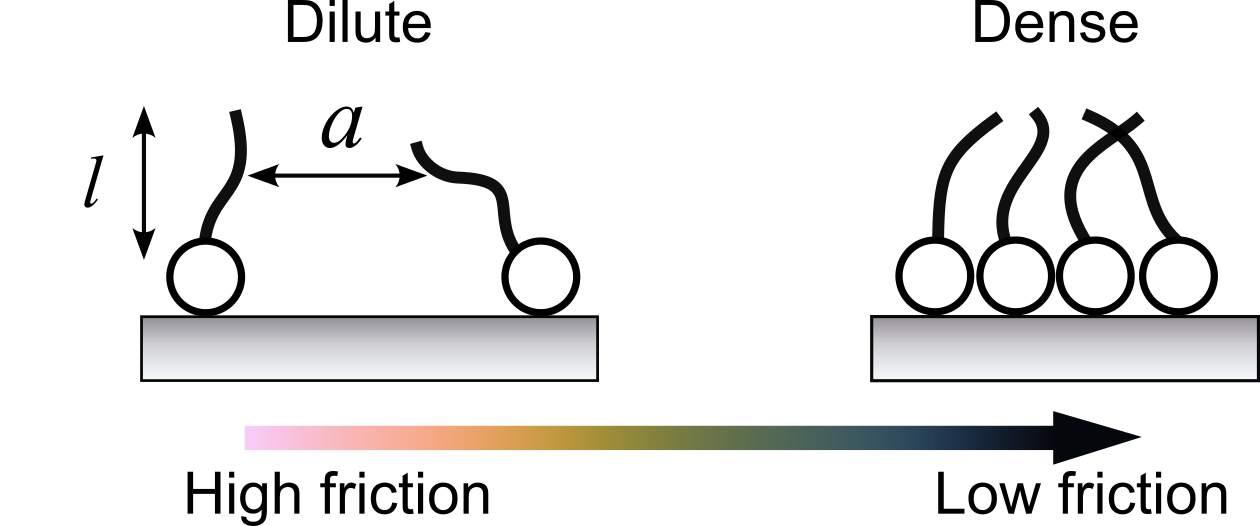}
    \caption{Schematic of the charge-mediated arrangement of surfactant molecules on the surface. Dense packing (small $a$) and long alkyl chains (large $\ell$) promote effective lubrication.}
    \label{fig:arrange}
\end{figure}

In this picture, the friction reduction sets in once the surfactant surface coverage becomes sufficient. This is confirmed by our observation that increasing the bulk surfactant concentration sharply reduces the coefficient of friction, followed by saturation (see Fig.~\ref{fig:setup}(c), CPC surfactant). The coverage at equilibrium follows the Langmuir adsorption isotherm~\cite{penfold2007surfactant}, $\Gamma = \Gamma_{m} [c K/ (1+ cK)]$, where $\Gamma_m$ is the maximum surface coverage and $K$ is the adsorption constant. This allows quantitative reconstruction of the surfactant surface density; see details in Fig.~S6. At low concentration (e.g., $c = 0.035$ mM), the surface is sparsely covered by the surfactant, $\sim$0.42 molecule/nm$^2$. The interactions between alkyl tails are weak, giving a picture that the adsorbed molecules may tilt or lie flat on the surface (Fig.~\ref{fig:arrange}-left). The lubricant layer is mechanically fragile and prone to collapse or be washed away under shear. Yet, a tenfold increase in concentration increases the surface coverage to $\sim$2 molecules/nm$^2$; the dense packing forces alkyl tails into brush-like conformations~\cite{milner1991polymer} (Fig.~\ref{fig:arrange}-right), leading to effective and robust lubrication.

Interchain interactions within the surfactant layer depend strongly on chain length $\ell$. For the short alkyl chains, weak van der Waals stabilization ($\sim$7 kJ/mol per \ch{CH2}~\cite{carpick1997scratching}) yields energetically fragile layers with low resistance to asperity penetration~\cite{silbert2014normal}. By contrast, long chains such as C$_{16}$ provide cumulative stabilization exceeding 100 kJ/mol, which is comparable to the energy of a covalent bond (100–300 kJ/mol). This additionally reinforces interdigitation and produces a cohesive, quasi-solid film that resists squeeze-out and shear. 
\textit{Conclusion---}We have established that friction between hard surfaces can be dramatically reduced by a universal charge-matching principle: surfactants of the opposite charge as the solid surface self-assemble into robust molecular brushes that sustain lubrication beyond conventional limits, even under extreme pressures. Simple molecular tuning—by varying surface coverage and/or chain length—enables unprecedented control over macroscopic friction. This charge-mediated mechanism requires no surface modification and provides a general route to ‘smart’ lubricants with broad implications, from advanced manufacturing to biomedical applications~\cite{jani2025compressing,lin2021recent}.



\textit{Acknowledgments---}This work was supported by the European Research Council (ERC) under the European Union’s Horizon 2020 research and innovation program (Grant No. 833240) and the ERC program (n-AQUA, grant No. 101071937). K.X. gratefully acknowledges the funding support from the Marie Skłodowska-Curie Actions Postdoctoral Fellowship (Grant No. 101150851). The authors also thank Gertjan Bon for preparing the glass substrates and P. Kolpakov for surface tension measurements.




%

\end{document}


\title{Supporting Information:\newline On the Slipperiness of Surfactants: Charge-Mediated Friction Control at the
Molecular Scale}

\author{Kaili Xie$^{1}$, Julie Jagielka$^{1}$, Liang Peng$^{1}$, Yu Han$^{2}$, Yedam Lee$^{2}$, Steve Franklin$^{1,3}$,
Yongkang Wang$^{2}$, Arsh Hazrah$^{2}$, Mischa Bonn$^{1,2}$, Joshua Dijksman$^{1}$, and Daniel Bonn$^{1}$ }

\maketitle 

\begin{affiliations}
  \item 
   van der Waals-Zeeman Institute, Institute of Physics, University of Amsterdam, Amsterdam, The Netherlands

   \item 
   Department of Molecular Spectroscopy, Max Planck Institute for Polymer Research, 55128, Mainz, Germany

   \item 
   Advanced Research Center for Nanolithography, 1098XG, Amsterdam, The Netherlands

    $\ast$ \textit{Corresponding author:\\ Daniel Bonn: d.bonn@uva.nl }
\end{affiliations}

\cleardoublepage
\section{Materials and experimental protocols}
Cationic surfactants, cetylpyridinium chloride (CPC, 99\%) and  n-alkyltrimethylammonium bromide (C$_n$TAB, n = 6, 8, 10, 12, 16) were obtained from Sigma-Aldrich. 
Anionic surfactants, bis(2-ethylhexyl) sulfosuccinate sodium salt (AOT, 98\%) and sodium dodecyl benzene sulfonate (SDBS, technical grade) were from Sigma-Aldrich. Non-ionic surfactant Triton X-100 (laboratory grade) was also obtained from Sigma-Aldrich. Sodium chloride ($\geq$99.5\%) was purchased from Carl Roth GmbH \& Co. KG. Deuterium oxide (\ch{D2O}, for NMR, 99.8 atom\% D) was purchased from Thermo Scientific Acros. All surfactants were first prepared in Milli-Q water (18.2 M$\Omega\cdot$cm, Millipore) at high concentration as stock solutions, and subsequently diluted with glycerol (from VWR) and water to the desired concentration, yielding a final composition of 75 wt\% glycerol and 25 wt\% water. All solutions were thoroughly mixed and stored in 40 mL clean glass vials sealed with Teflon-lined caps (from Sigma-Aldrich) to prevent concentration changes due to evaporation. 



Prior to friction measurements, all spheres and substrates were thoroughly cleaned by sonication for 15 minutes in each of the following solvents: acetone, ethanol, and isopropyl alcohol (VWR), followed by rinsing with Milli-Q water and drying under a nitrogen stream. For \ch{CaF2} surfaces, a plasma treatment of 20 minutes was applied before frictional experiments. This ensures a positively charged surface in an aqueous environment. 
\bigskip

\begin{figure}[h!]
    \centering
    \includegraphics[width=0.8\linewidth]{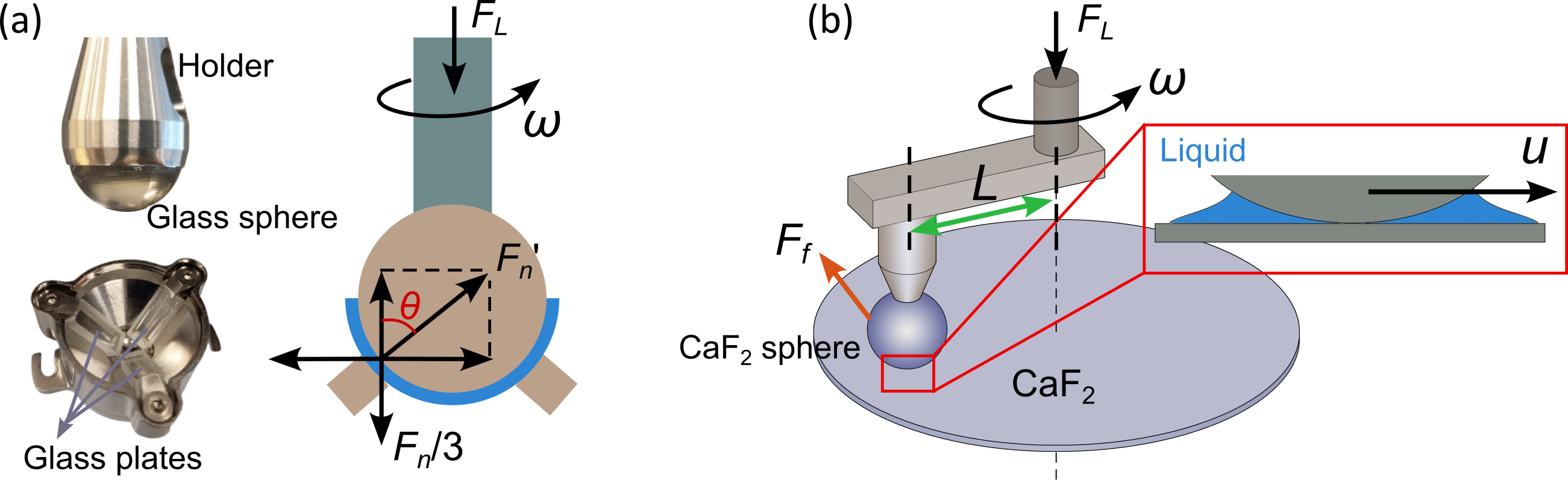}
    \caption{Experimental setups for frictional measurements. (a) Anton Paar tribometer. In measurements, the glass sphere contacts with three glass plates immersed in the liquid. The glass plates were inclined at $\theta$ = $\pi/4$. (b) Custom-built tribological setup. The 4 mm diameter of \ch{CaF2} sphere was fixed on a cantilever that is equipped with a rheometer head. The distance $L$ = 10 mm is defined between the center of the sphere and the centerline of the rotation shaft. The close-up image shows that the interface is completely immersed in the liquid. In both systems, the frictional forces were deduced from the measured torque.   }
    \label{fig:SI_setup}
\end{figure}

For the commercial tribometer, a glass ball (12.7 mm diameter, soda lime, from Anton Paar) was mounted onto a measuring geometry (BC 12.7, 48837), while three soda lime glass plates (15.3 mm $\times$ 6.2 mm $\times$ 3.9 mm) were fixed inside a reservoir (T-PTD 200), see the picture in Figure S\ref{fig:SI_setup}(a). A volume of 1 mL of solution was added in all experimental tests. Prior to each measurement, the system was allowed to equilibrate for about 5-10 minutes. A cover was applied to minimize evaporation.

To measure friction between sliding \ch{CaF2} surfaces, we used a customized setup (Figure S\ref{fig:SI_setup}(b)) described earlier~\cite{peng2022nonmonotonic}, since the \ch{CaF2} sphere is too small to fit onto the commercial tribometer. The \ch{CaF2} sphere has a diameter of 4 mm (from Edmund Optics), mounted on a holder connected to a rheometer geometry. To roughen the surface, the \ch{CaF2} sphere was sandblasted before use. The bottom \ch{CaF2} plate (diameter 25 mm, thickness 2 mm, from Hangzhou Shalom Electro-optics Technology) was fixed on a horizontal platform. 
A rheometer head (DSR 502, Anton Paar) was used to precisely control the sliding velocity while simultaneously measuring the torque. Before applying a normal load, a volume of \SI{200}{\micro\liter} solution was introduced between the sphere and the substrate, ensuring that the interface was fully immersed in the liquid environment.
\bigskip

\begin{figure}[h!]
    \centering
    \includegraphics[width=0.7\linewidth]{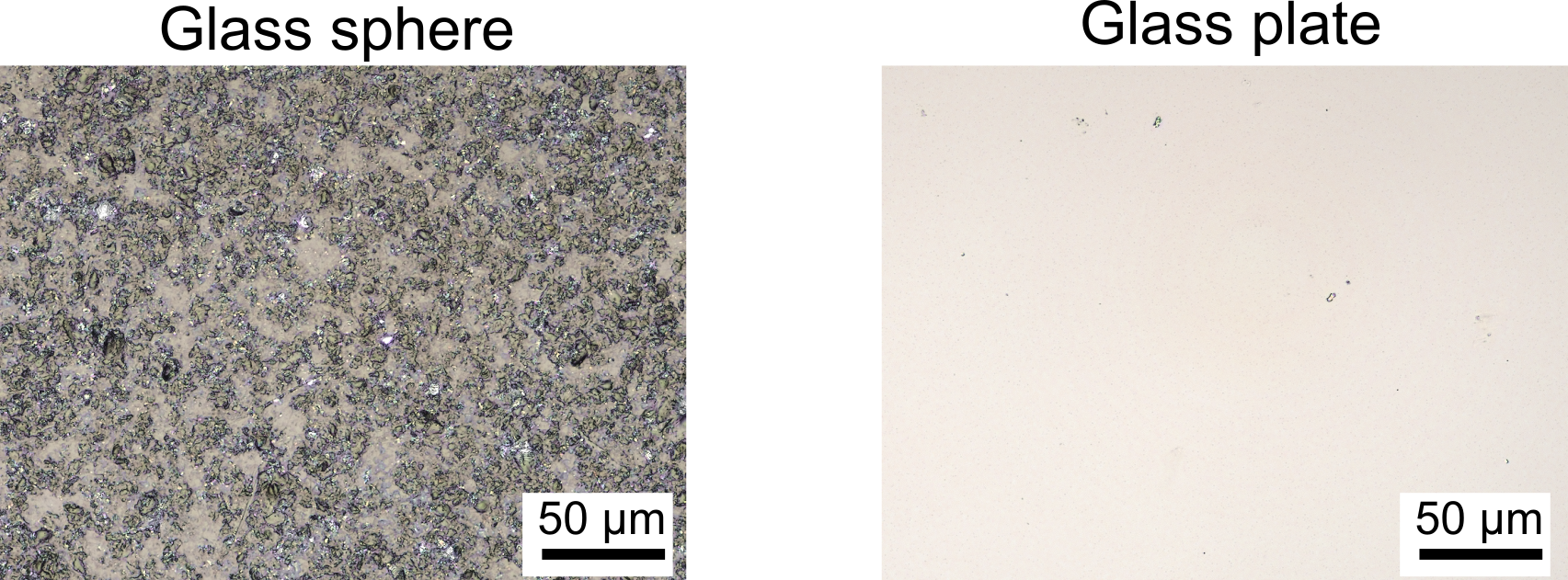}
    \caption{Topography of the glass sphere and plate captured by a laser profilometer.  }
    \label{fig:SI_glassRoughness}
\end{figure}

\section{Friction measurement}
We measured the friction of sliding glass-glass surfaces using the commercial tribometer (Figure S\ref{fig:SI_setup}(a)). The upper glass sphere has a surface roughness, $Sq = 425 \pm 60$ nm, which is much larger than the bottom glass plate ($Sq = 6 \pm 2$ nm), see Figure S\ref{fig:SI_glassRoughness}. The surface roughness was determined by a laser profilometer (Keyence laser microscope, VK-X3050).

\begin{figure}[h!]
    \centering
    \includegraphics[width=0.8\linewidth]{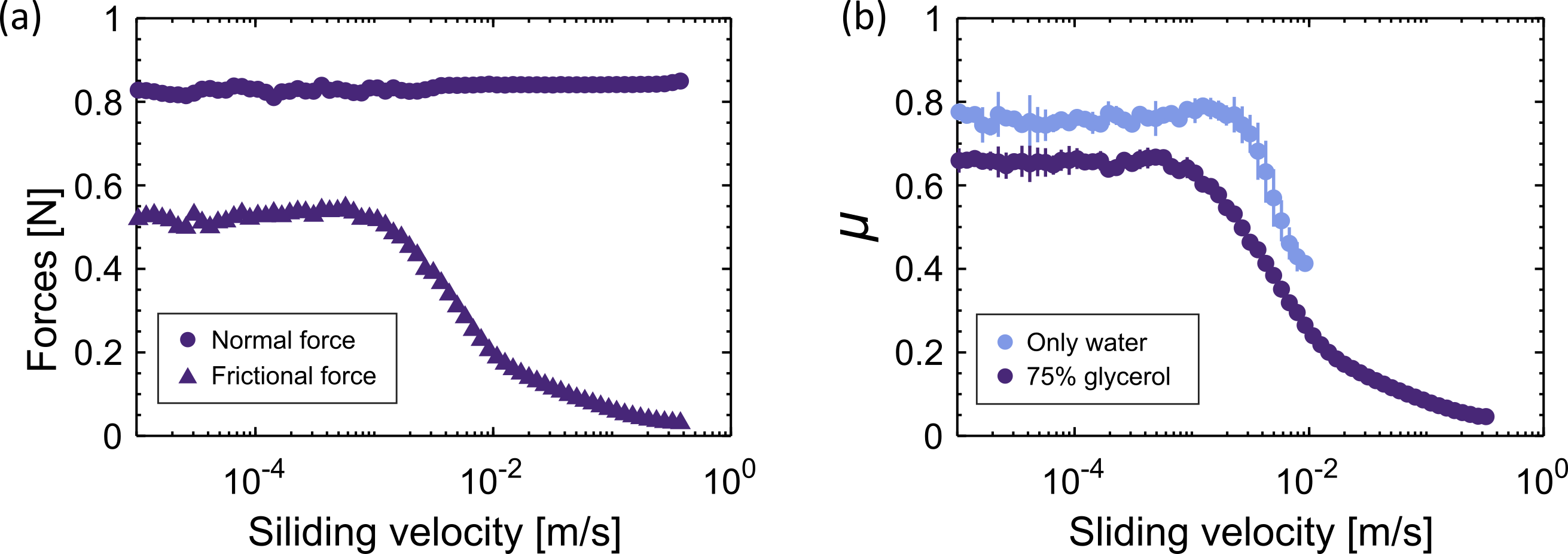}
    \caption{Frictional measurement using tribometer. (a) The normal and frictional forces versus sliding velocity at one of the sliding glass interfaces. (b) The CoF as a function of sliding velocity using only water and glycerol-water mixture as lubricant. In water, the measurement stopped at a velocity of 10$^{-2}$ m/s due to the appearance of a shrill noise. While this does not happen when adding 75 wt\% glycerol.}
    \label{fig:SI_forces}
\end{figure}

During measurement, a constant normal load was applied by the rheometer head, while the sliding velocity was varied in a few orders of magnitude (Figure S\ref{fig:SI_forces}). 
The total friction force can be calculated from the torque $M$ measured by the rheometer, 
\begin{equation}
    \centering
    F_f = \frac{M}{R \sin{\theta}},
    \label{eq:friction_f}
\end{equation}
where $R$ is the radius of the glass sphere and $\theta$ is the inclination angle of the glass plate, which is $\pi/4$. Ideally, this frictional force is related to the normal forces applied to the three sliding interfaces with a mean coefficient of friction (CoF) defined as $\mu$,
\begin{equation}
    \centering
    \mu \left(   F_{n1} + F_{n2} + F_{n3} \right) = F_f,
    \label{eq:friction_nr}
\end{equation}
where $F_{n1}$, $F_{n2}$, and $F_{n3}$ are the normal forces that are applied to each glass-glass interface. From the geometry, it is known that,
\begin{equation}
    F_{n1} = F_{L1} / \cos{\theta}, F_{n2} = F_{L2} / \cos{\theta}, F_{n3} = F_{L3} / \cos{\theta}.
    \label{eq:normal_f}
\end{equation}
The total load $F_L$ that the rheometer head applied satisfies, 
\begin{equation}
    \centering
    F_{L1} + F_{L2} + F_{L3} = F_L.
    \label{eq:load}
\end{equation}
Combining Equations (\ref{eq:friction_f})-(\ref{eq:load}), we can calculate the mean CoF,
\begin{equation}
    \centering
    \mu = \frac{M}{R F_L \tan{\theta}}.
    \label{eq:meanCoF}
\end{equation}
Here, we note that the CoF measurement is independent of the number of glass–glass contact points, even though, at high sliding velocities, not all three contact points may be maintained.
\bigskip

\bigskip
\bigskip

\begin{figure}[h!]
    \centering
    \includegraphics[width=0.4\linewidth]{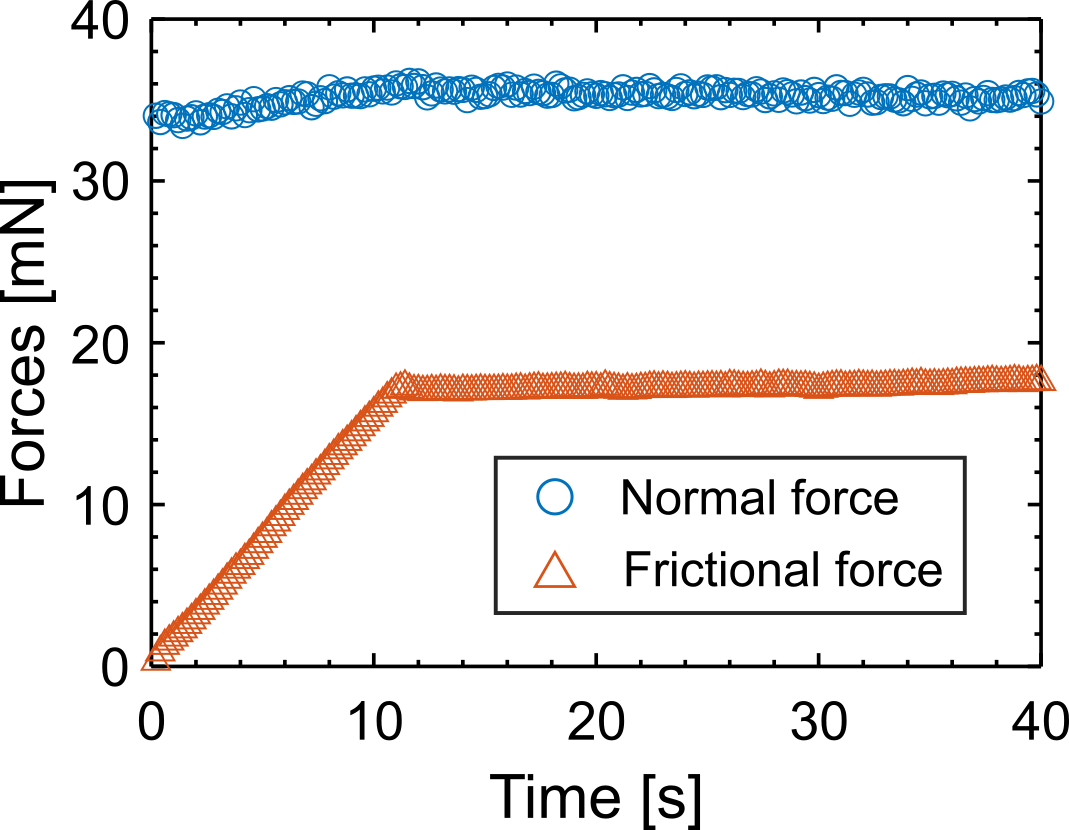}
    \caption{The normal and frictional forces versus time in frictional measurement using our custom-built tribological setup.  }
    \label{fig:SI_forcetime}
\end{figure}

Figure S\ref{fig:SI_forces}(a) shows that the normal force applied to each sliding surface was maintained constant across various sliding velocities. The plateau in the frictional force demonstrates the boundary lubrication regime. Figure S\ref{fig:SI_forces}(b) shows the comparison of the CoF with and without the addition of glycerol to water. When using only water, the system generates shrill noise and vibrations at sliding velocities above 10$^{-2}$ m/s, at which the measurement has to be stopped. Adding 75 wt\% glycerol in water, the CoF slightly decreases as glycerol acts as a better lubricant. The measurement can be maintained up to a sliding velocity of 0.5 m/s, covering both the boundary and mixed lubrication regimes.  
Therefore, in all our experiments, we used a 75 wt\% glycerol and 25 wt\% water mixture as the solvent, in which various types and concentrations of surfactants were dissolved.
Control experiment contains no surfactant.

We measured the friction at the \ch{CaF2}-\ch{CaF2} interfaces using the customized tribological setup (Figure S\ref{fig:SI_setup}(b)). The frictional force is calculated as $F_f = M/L$, where $M$ is the torque and $L$ is the distance between the center of the sphere and the centerline of the rotation shaft. The CoF is determined from the plateau where the frictional force becomes time-independent (Figure S\ref{fig:SI_forcetime}).

\section{Solvent viscosity at different concentrations of the surfactant}

\begin{figure}[h!]
    \centering
    \includegraphics[width=0.4\linewidth]{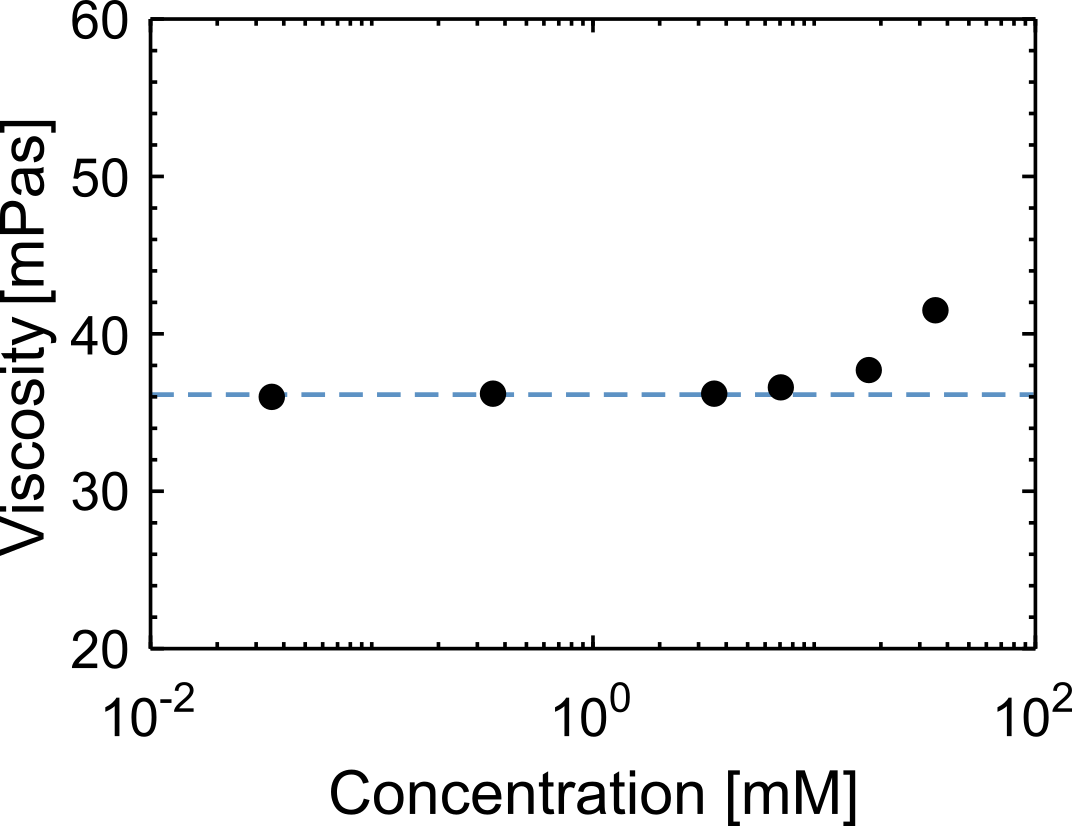}
    \caption{Viscosity of 75 wt\% glycerol-25 wt\% water mixture with different concentrations of CPC surfactant. The dashed line indicates the viscosity of the solvent without surfactant.}
    \label{fig:SI_eta}
\end{figure}

\section{UV–vis spectroscopy measurement}
\begin{figure}[h!]
    \centering
    \includegraphics[width= 0.8\linewidth]{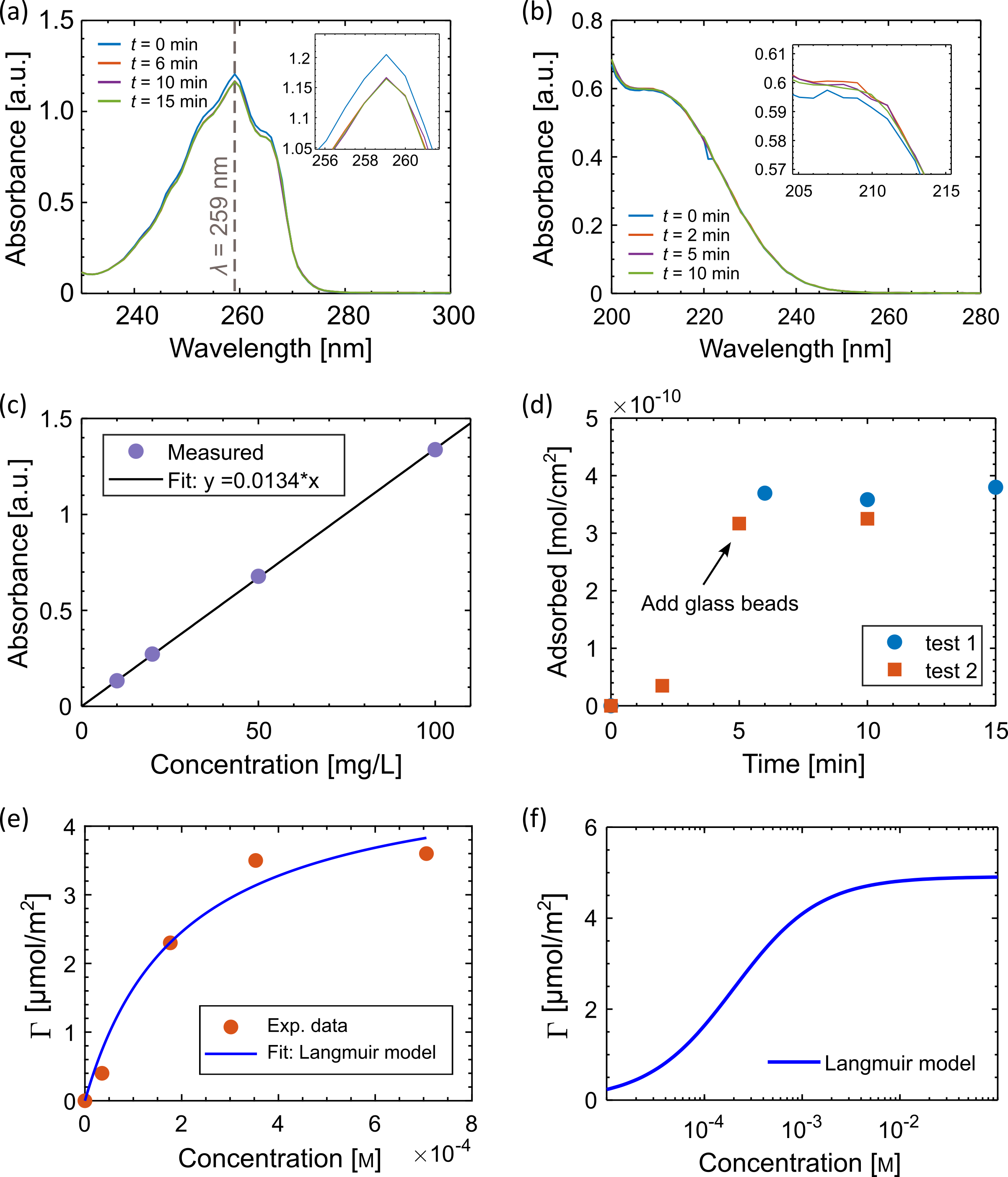}
    \caption{UV-vis spectroscopic measurement. (a) Absorbance spectra of a 0.01\% CPC solution at different wavelengths. The inset shows a close-up of the absorbance at maxima where the wavelength $\lambda = $ 259 nm. (b) Absorbance spectra of 0.1\% AOT solution. The inset shows that the absorbance intensity roughly has no change. $t$ = 0 min indicates the absorbance without adding glass beads. Others are with glass beads. (c) Calibration of absorbance of CPC solution at different concentrations. (d) The amount of adsorbed CPC molecules per unit area onto the glass surface. Two independent experiment tests were performed.  (e)  Surface excess of CPC surfactant on the glass at equilibrium for different concentrations. The solid line shows the fit using Langmuir model (Equation \ref{eq:langmuirModel}). (f) Reconstructed adsorption isoterms }
    \label{figSI:uv_vis}
\end{figure}

The adsorption of the cationic surfactant (CPC) and the anionic surfactant (AOT) onto the glass surface was measured using a UV–vis spectrophotometer (Cary 50), see Figure S\ref{figSI:uv_vis}. 
Briefly, a volume of 3 mL water was first placed in a quartz cuvette ($l$ = 1 cm, path length) and used as the baseline. The surfactant solution (same volume) was scanned over a wavelength range of 200–300 nm ($t$ = 0 min). Afterwards, a certain amount of glass beads (nominal diameter 0.8 mm) was added to the cuvette containing the surfactant solution. The absorbance of the supernatant solution was subsequently measured. Figure S\ref{figSI:uv_vis}(a) shows that the absorbance of the CPC solution at $\lambda$ = 259 nm decreases and reaches saturation when the glass surfaces are fully covered by the surfactant. In contrast, the absorbance of the AOT solution $\lambda$ = 210 nm shows negligible change (Figure S\ref{figSI:uv_vis}(b)).

The relationship between absorbance and surfactant concentration is given by the Beer–Lambert law,
\begin{equation}
    \centering
    A = \varepsilon l c,
    \label{eq:BeerLambert}
\end{equation}
where constant $\varepsilon$ is the molar extinction coefficient and $c$ is the concentration. By calibrating the absorbance of CPC at different concentrations, $\varepsilon$ can be obtained (Figure S\ref{figSI:uv_vis}(c)).

From Equation \ref{eq:BeerLambert} and the calibration curve, the adsorbed CPC surfactant on the glass surface at equilibrium is shown in Figure S\ref{figSI:uv_vis}(d). We thus can obtain the  adsorption isotherm of CPC surfactant at various concentrations, shown in Figure S\ref{figSI:uv_vis}(e). 
The amount of CPC adsorption on the glass surface increases as initial CPC concentration increases to 0.35 mM, while it roughly becomes constant with a higher concentration. This clearly indicates that the adsorption process can be described by the Langmuir isotherm model~\cite{penfold2007surfactant}, 
\begin{equation}
    \centering
    \Gamma = \Gamma_{m}  \frac{K c}{1 + Kc},
    \label{eq:langmuirModel}
\end{equation}
with a maximum adsorption $\Gamma_m$ and an adsorption equilibrium constant $K$. By fitting our experimental data using Equation \ref{eq:langmuirModel}, we thus reconstruct the adsorption isotherm in Figure S\ref{figSI:uv_vis}(f), where $\Gamma_m$ = 4.91 $\mu$mol/m$^2$ and $K = 5 \times 10^3$ L/mol.

\section{HD-SFG Methods}
\subsection{\textbf{Sample Preparation}} 
NaCl salt was baked in an oven at 650\,\textdegree C for 8 hours before use. NaCl solutions were prepared by dissolving the required amount of NaCl in Milli-Q water. Solutions containing CPC or AOT surfactant were prepared by first dissolving the appropriate amounts of 0.1\% stock solution and NaCl into clean vials to reach final NaCl concentrations of 0.2 mM, 5 mM, and 100 mM. Milli-Q water was then added to bring the total solution mass to 40 g. The vials were sealed and mixed on a roller until clear (approximately 1 hour).

Glass substrates (15 mm × 10 mm × 1 mm)  were cleaned sequentially with acetone, 2-propanol, ethanol, and Milli-Q water in an ultrasonic bath for 5 minutes each. CaF\textsubscript{2} substrates (25 mm diameter, 2 mm thickness) were rinsed sequentially with acetone, 2-propanol, ethanol, and Milli-Q water. After cleaning, two stripes of 100 nm thick gold coating were thermally evaporated onto the substrates through a shadow mask to serve as phase references. Before measurements, the substrates were cleaned again with acetone, 2-propanol, ethanol, and Milli-Q water and dried using nitrogen flow.

\subsection{\textbf{HD-SFG Measurement}}
The HD-SFG measurements were carried out with a non-collinear HD-SFG setup. This setup was constructed using a Ti:Sapphire regenerative amplifier laser system (Spitfire Ace, Spectra-Physics,  800 nm, 40 fs,  5 mJ, 1 kHz). Detailed description can be found in Ref.~\cite{seki2021direct,wang2023chemistry}. In this HD-SFG configuration, the visible (800 nm, 6 µJ), local oscillator (LO), and infrared (2800-3800 cm\textsuperscript{-1}, 3 µJ) lights impinge non-collinearly from the optically transparent glass substrate, passing through the glass to overlap at the glass-water interface. The reflected LO light interferes with the sum-frequency signals generated from the water in the ‘reflected’ direction, producing a heterodyned sum-frequency output that enables access to the Im$\chi^{(2)}$ signals. 

The HD-SFG spectra were collected under the \textit{ssp} polarization combination (SFG: \textit{s}-, visible: \textit{s}-, IR: \textit{p}-polarized). Each substrate-water spectrum was acquired with an exposure time of 10 minutes × 5 times, while spectra from the substrate-gold interface were recorded with an exposure time of 1 minute × 5 times. To determine the phase of the gold, we followed the protocol in Ref.~\cite{wang2023chemistry}. Specifically, we measured the Im$\chi^{(2)}$ spectra of the substrate-\ch{D2O} interface by normalizing its HD-SFG signal to that of the substrate-gold interface, assuming that the Im$\chi^{(2)}$ spectrum of the \ch{D2O} interface is zero across the O-H stretching region. A height displacement sensor (CL-3000, Keyence) was used to monitor the sample height during phase measurements. Based on the Gouy-Chapman model and following previously established protocols~\cite{seki2021direct,wang2023chemistry,wen2016unveiling}, the surface charge densities were obtained with the differential Im$\chi^{(2)}$ spectra of SFG at three different concentrations of NaCl: 0.2 mM, 5 mM, and 100 mM.

\section{Statistical Analysis}
The friction measurements in tribology presented in this work were averaged from 2–4 independent experiments. The error bars show the standard deviation.

\section{References}   
